# Which US and European Higher Education Institutions are visible in ResearchGate and what affects their RG Score?

Benedetto Lepori, Faculty of Communication Sciences, Università della Svizzera italiana, via Buffi 13, CH-6904 Lugano, Switzerland, blepori@usi.ch (corresponding author).

Michael Thelwall, Statistical Cybermetrics Research Group, School of Mathematics and Computer Science, University of Wolverhampton, Wolverhampton, United Kingdom, M.Thelwall@wlv.ac.uk.

Bareerah Hafeez Hoorani, Faculty of Communication Sciences, Università della Svizzera italiana, via Buffi 13, CH-6904 Lugano, Switzerland, bareerah.hafeez.hoorani@usi.ch.

## Abstract

While ResearchGate has become the most popular academic social networking site in terms of regular users, not all institutions have joined and the scores it assigns to academics and institutions are controversial. This paper assesses the presence in ResearchGate of higher education institutions in Europe and the US in 2017, and the extent to which institutional ResearchGate Scores reflect institutional academic impact. Most of the 2,258 European and 4,355 US higher educational institutions included in the sample had an institutional ResearchGate profile, with near universal coverage for PhD-awarding institutions found in the Web of Science (WoS). For non-PhD awarding institutions that did not publish, size (number of staff members) was most associated with presence in ResearchGate. For PhD-awarding institutions in WoS, presence in RG was strongly related to the number of WoS publications. In conclusion, a) institutional RG scores reflect research volume more than visibility and b) this indicator is highly correlated to the number of WoS publications. Hence, the value of RG Scores for institutional comparisons is limited.

# 1 Introduction

The academic social networking site ResearchGate has become an important component of the scholarly communication landscape. Between its foundation in 2008 and 2017, it claimed to have attracted 14 million members[1], although recruitment may now be stabilising (Ortega, 2017). If most ResearchGate members are academics or doctoral students, this accounts for a substantial portion of the world's practicing academic researchers. According to a Nature survey, 48% of science and engineering researchers and 35% of social science, arts and humanities scholars visit ResearchGate regularly, which is five times more than its nearest competitor, the academic social networking site Academia.edu (Van Noorden, 2014). Members can use their ResearchGate profile to showcase their activities and publications, whereas institutional profiles aggregate the achievements of affiliated researchers. The most controversial aspect of ResearchGate is the scores that it prominently displays for academics, publications and institutions. These currently include the number of reads, citations and recommendations, as well as the flagship RG Score. The prominence of the RG Score encourages researchers to take it seriously, even though ResearchGate does not explain what it means and how it is calculated. A third of academic ResearchGate users pay attention to its metrics (question 4 in the figure of: Van Noorden, 2014), and most academics in a European survey thought that scores from such systems would be increasingly important (Jamali *et al*, 2015). At the institutional level, the RG Score may be used by students and junior researchers when selecting institutions to attend, and even by research managers seeking an easy source of ranking information for policy and promotional purposes (Wilsdon *et al*, 2015).

ResearchGate's headline description for their RG Score is "The RG Score is a metric that measures scientific reputation based on how all of your research is received by your peers."[2] It states that contributions, interactions and reputation all help (listed in this order), but the headline also includes the statement "Interactions form the basis of your RG Score"[3] in the middle of the page. It is not clear from this statement whether interactions *within ResearchGate* are more important, or how those different components are measured or balanced. The lack of transparency of the RG Score is an important issue because its prominent placement on the site means that busy researchers or students may interpret it at face value, leading to incorrect decisions (Martín-Martín *et al*, 2016).

Despite several investigations of ResearchGate, it is not clear how universal its uptake is, whether there are substantial islands of non-users (e.g., countries), as well as the extent to which it has penetrated non-research higher education institutions, such as colleges. Previous research, reviewed below, suggests that RG Scores for individual academics primarily reflects their level of activity within the site. If this holds, RG Scores at the institutional level would not reflect traditional scholarly impact to an extent that would make them useful reputation indicators, but rather the volume of research activity. However, this is contradicted by early analyses (see below), which investigated older versions of the RG Score that included impact factors. Moreover, none of these studies controlled for the cofounding effect of institutional size, which tends to be correlated with academic reputation. It is therefore not clear whether current institutional RG Scores are reasonable indicators of institutional academic impact.

In response, this article analyses presence and institutional RG Scores in ResearchGate for 6,613 Higher Educational Institutions (HEIs) in Europe and the US, a sample that nearly comprises the entire population of HEIs awarding degrees at least at the bachelor level and includes all 84 US and European universities within the first 100 universities in the 2013 edition of the Shanghai ranking. A variety of factors that might affect presence and institutional RG Scores are controlled for, including institutional size, PhD-awarding capability, the number of publications in the Web of Science and their mean field normalized citation count. This paper therefore analyses the largest sample yet from two important geographical locations for science in conjunction with the largest set of covariates to offer the most systematic quantitative investigation into the reasons behind both ResearchGate's increased adoption and the RG Score at the institutional level.

---

[1] www.researchgate.net/aboutus.AboutUs.html, last visited 05.12.2017.

[2] https://www.researchgate.net/RGScore/FAQ last visited 13.12.2017.

[3] https://www.researchgate.net/RGScore/FAQ last visited 13.12.2017.

## 2 Background

ResearchGate is an academic social networking site in the sense that it targets academics and has academic-specific features (publication uploading, citation indexing), as well as traditional social networking functions. ResearchGate is like Academia.edu in this regard (Ovadia, 2014), but with more regular visitors in 2014 (Van Noorden, 2014) and more visitors reported by Alexa.com in May 2018 (global rank 246 for ResearchGate and 624 for Academia.edu).

### 2.1 Scholarly communication

In addition to providing a platform for scientific work, view/download counts and the RG Score, the social networking features incorporated by ResearchGate include the ability to privately message other members, to publicly connect to them, and to share in public discussions. In theory, members may take advantage of all, some or none of these features. They may use ResearchGate as their primary communication and dissemination tool, or employ it alongside more traditional strategies, such as publishing in journals and presenting at conferences.

A survey of scientists found ResearchGate to be visited regularly by almost half of all academics, ahead of all other academic-related and free online sites, except Google Scholar (Van Noorden, 2014). Its dominance was confirmed for a sample of Norwegian researchers, except in the humanities (Mikki *et al*, 2015), however successful researchers seem less likely to be active users of social web sites (Mas-Bleda *et al*, 2014). Joining and using an academic social network site is an investment in time (Ovadia, 2014) and so these activities may reflect a belief that the social or reputational benefits outweigh the time investment (Williams and Woodacre, 2016).

In some instances, students have also joined ResearchGate in large numbers. In Spain, both undergraduates and postgraduates are represented as members of universities (Iglesias-García *et al*, 2017).

Academics seem to regard academic social network sites as an important part of their professional identity or as a repository for their publications (Corvello, Genovese, & Verteramo, 2014; Jordan, 2017). An online convenience sample survey of the factors associated with ResearchGate's (and Academia.edu) adoption among 6,139 Italian scholars found that the participants mainly used the site to increase the visibility of their academic output (Manca and Ranieri, 2017). Unlike Twitter, academic social networks do not seem to be used for informal personal interactions (Jordan, 2017). An analysis of 55 Swiss management researchers found that engaging with the platform helped to attract followers to the site, as did academic seniority and the impact of publications (Hoffmann *et al*, 2016). ResearchGate is perceived by some members to be ineffective (45% in one convenience sample survey), and the ability to compare scores between researchers can be a source of stress (15%) (Muscanell *et al*, 2017). In some cases it can also be viewed as a time-wasting activity (Madhusudhan, 2012), while others regard it as being helpful for publication sharing (15%) and networking (25%) (Muscanell *et al*, 2017; see also Meishar-Tal and Pieterse, 2017). Academics may also use ResearchGate to form study groups (Chakraborty, 2012).

An important aspect of social networking is the ability to get answers to questions that are difficult to address through more formal methods, such as literature reviews. The Question and Answer (Q&A) section of ResearchGate may therefore play a role in scholarly communication, especially for academics that are not embedded within large research-active departments. One study found that fast, long, focused answers from authoritative researchers tended to attract positive ratings on the ResearchGate Q&A site, suggesting that it tends to play an informational (rather than purely social) role (Li *et al*, 2015). This contrasts with a study of ResearchGate data from 2013, which found that most questions on the Q&A site went unanswered (Alheyasat, 2015), but this may have subsequently changed. There seems to have been a tendency for contributions to become more serious and less social over time, perhaps as the Q&A feature became recognised as a valuable tool (Goodwin *et al*, 2014). Overall, questions can be requests for specific information or a discussion about a topic. Responses can be highly detailed, with rich content that demonstrates a substantial attempt to give an effective response or demonstrate knowledge (Jeng *et al*, 2017).

## 2.2   Publication-level indicators

In mid-2017, ResearchGate had found, or permitted researchers to upload, a large corpus of academic texts, with half of the non-open access publications breeching publisher copyright (Jamali *et al*, 2015), and with disciplinary variations (Thelwall and Kousha, 2017a). It indexed the citations from these publications, allowing it to report citation counts for articles in researcher profiles, which gave the site an advantage over Scopus and WoS in finding early citations from preprints (Thelwall and Kousha, 2017b). An investigation into publications published in 2014 by top Spanish universities found them to be mostly (55%) uploaded to ResearchGate but rarely (11%) to institutional repositories (Borrego, 2017). ResearchGate is also the single largest source of open access full text publications, at least according to Google Scholar (Mikki *et al*, 2015; Jamali and Nabavi, 2015; see also: Laakso *et al*, 2017). Uptake may be lower in some parts of the world, including Africa (Baro and Eze, 2017).

ResearchGate's ability to index academic papers seemed to be under threat in late 2017 due to publisher lawsuits (Chawla, 2017). In response ResearchGate removed 1.7 million papers from five different publishers in November of 2017, however it is not clear whether this affected subsequent RG Scores.

The download and view counts provided for publications by ResearchGate are alternative impact indicators that have low or moderate correlations with citation counts, depending on discipline. They probably reflect a wider audience than for citations, such as from students in some fields (Thelwall and Kousha, 2017a). Correlations with citation or quality-related indicators are a useful way to investigate new indicators because statistically significant correlations provide evidence that the new indicator is not random and is related to scholarly impact in some way (Sud and Thelwall, 2014). If the numbers involved are not small (Thelwall, 2016), then it also gives evidence of the strength of the relationship between scholarly impact and the new indicator.

## 2.3   The RG Score for individual researchers

RG Scores draw users' attention by being prominently displayed near the top of each researcher's profile (Kraker and Lex, 2015). An early analysis suggested that journal impact factors were important for the RG Scores of individual researchers (Jordan, 2015), but ResearchGate subsequently removed impact factor related information from the site and presumably also removed it from the RG Score. An investigation of Spanish research council scientists confirmed that ResearchGate is more popular than Academia.edu and found a moderate correlation between RG Score and Google Scholar citations. This may have been due to disciplinary differences in citation cultures, in conjunction with RG Scores, while probably including impact factors at the time – so high citation specialisms would have high Google Scholar citation counts and high RG Scores (Ortega, 2015).

A more recent analysis of RG scores for individual researchers from August 2016 found that extensive activity within ResearchGate was essential for obtaining a high score. Publications alone could generate a moderately high score, but not as high as when answering questions on the site (Orduna-Malea *et al*, 2017). Uploading full-text versions of papers can also help RG Scores (Copiello and Bonifaci, 2018).

Investigations that have correlated citation counts with RG Scores for individual academics have produced mixed results. For communication sciences and disorder research scholars in the US and Canada, RG Scores had a strong correlation (>.44; the exact figure is not reported) with Scopus h-indexes (Stuart *et al*, 2017). An investigation of bibliometricians found RG Scores to correlate very highly (0.9) with Google Scholar citations and the Google Scholar h-index, and highly (0.6) with the number of ResearcherID publications (Martín-Martín *et al*, 2018). In both of these cases, the magnitude of the correlation will be affected by the samples, including both junior and senior researchers. In contrast, an investigation of top management scholars in Taiwan found no relationship between RG Scores and performance (Kuo *et al*, 2016). Thus, the current RG Score seems to primarily reflect activity within the site or full text uploading, rather than external reputation or achievements (Orduna-Malea *et al*, 2017).

## 2.4  The RG Score for institutions

The institutional-level RG Score is presumably compiled from the data used for institutional members. Larger universities may therefore tend to have higher scores by drawing upon a greater pool of data and, therefore, it is expected that institutional RG Scores correlate with size.

At the institutional level, aggregate RG downloads and view counts have low correlations (Spearman 0.2-0.3) with ad-hoc university rankings (THE ranking, QS, ARWU) and a correlation of 0.0 with Leiden citation-based university rankings (Thelwall and Kousha, 2015). Also at the institutional level, RG Scores had low correlations (0.2) with ad-hoc university rankings and small negative correlations (-0.1) with the Leiden citation-based rankings (Thelwall and Kousha, 2015). Stronger correlations were obtained between institutional RG Scores and average research quality for UK universities (Pearson 0.4) (Yu *et al*, 2016). The stronger correlation may be due to size correlating with quality for UK universities, giving a substantial spurious association, or the use of an inappropriate correlation coefficient for skewed data. The same dataset gave a Pearson correlation of 0.2 between institutional RG Scores and average citations per publication (not field normalised), which may also be affected by institutional size. An extremely high correlation was found for South African universities between institutional RG Scores and WoS (total) citations (0.97, n=23), presumably affected by size (Onyancha, 2015). These studies all used RG Scores when they may have included impact factors. A more recent study found no correlation between institutional RG Scores and QS rankings for Pakistani Universities (Ali *et al*, 2017). Finally, a recent study shows systematic differences when looking at participation in RG by academics of US universities (Yan *et al*, 2018). The study showed that universities with higher research activity levels have a higher proportion of active RG users that also have significantly higher RG scores. This study therefore suggests systematic differences in participation and RG visibility associated with the volume of institutional research activity.

# 3  Research questions

The following research questions address the adoption of ResearchGate and the meaning of RG Scores at the institutional level in US & Europe.

1. How comprehensive is ResearchGate membership amongst higher education institutions in the US and Europe, including those not awarding PhDs?
2. Which characteristics are most associated with institutional RG Scores for higher education institutions in the US and Europe: institutional size, publication output or academic reputation?

These questions are investigated with a sample of more than 6,500 HEIs in Europe and the US. It seems likely that uptake is lower in developing nations and that institutional RG Scores would be more variable, so it is useful to focus on a homogenous set of countries to investigate the impact of HEI characteristics on RG participation. The association between participation in RG and institutional RG Scores is tested with a range of institutional characteristics suggested by the literature, including organizational size (number of staff and students), the number of publications in Web of Science (WoS) and the publication impact in WoS. The results should provide insights into the potential significance of RG Scores for the purposes of institutional evaluation.

# 4 Methods

## 4.1 HEI sample

The sample for the study is derived from two reference datasets, the Integrated Postsecondary Education Data System for the US (IPEDS; http://nces.ed.gov/ipeds/) and the European Tertiary Education Register dataset (ETER; www.eter-project.com). ETER covers all EU-28 member states, EEA-EFTA countries (Iceland, Liechtenstein, Norway and Switzerland) and EU candidate countries (Former Yugoslav Republic of Macedonia, Montenegro, Serbia, Turkey). The French-speaking part of Belgium, Romania, Slovenia, Montenegro and Turkey were excluded because of missing data.

Both ETER and IPEDS provide broad coverage of institutions that deliver degrees at the tertiary level, corresponding to levels 5 to 8 of the International Standard Classification of Educational Degrees (ISCED; http://www.uis.unesco.org/Education/Pages/international-standard-classification-of-education.aspx).
IPEDS is a mandatory system for postsecondary institutions receiving federal aid in the US and therefore provides complete coverage, while ETER excludes small HEIs (below 200 students and 30 Full Time Equivalent [FTE] academic staff) and institutions delivering only professional degrees of less than three years (ISCED level 5). When compared with EUROSTAT data at the country level, coverage of student enrolments in the US is 100%, ETER coverage is 96% at ISCED 6-8, but only 52% at the ISCED 5 level (source: Eurostat statistics on tertiary education).

Both databases provide information on the institutional characteristics of HEIs, including budgets, staff, student enrolments and graduates. The dataset was supplemented with bibliometric data derived from the WoS copy maintained by CWTS in Leiden by searching the CWTS-WoS list of organizations for HEIs in ETER and IPEDS (Waltman *et al*, 2012). Both datasets were checked to match additional candidates, focusing on HEIs with many PhD degrees for which there was no match. Publication data were retrieved for 850 HEIs in ETER, which included 97.2% of the PhD degrees in the dataset, and for 410 HEIs in IPEDS, corresponding to 89.8% of PhD degrees. The lower coverage in the US is mostly due to two private distance universities with a high number of PhD students but few publications. There is no lower threshold for the number of publications (some HEIs identified in WoS have less than 10 publications in the dataset), but it is likely that not all HEIs with less than 100 publications in the reference year have been identified (particularly for the non-PhD awarding institutions). In other words, matching is extensive and includes almost all HEIs with large research volumes (the only exception being some medical research centres), but several HEIs with a low number of publications in WoS have probably not been identified.

The sample originally included 7,331 HEIs; after merging multi-campus HEIs with a single RG account (mostly in the US) and dropping a few problematic cases (see below), the final sample includes 6,613 HEIs (2,258 in Europe and 4,355 in the US). All institutional data are from 2013 (academic year 2013/2014). The time difference with ResearchGate data is not likely to be relevant for a cross-sectional analysis given the high stability of HEI data.

### 4.1.1 HEI variables

The following variables are derived from ETER and IPEDS at the HEI level.

- *The number of academic staff* (Full Time Equivalent). Both databases are based on working contracts; from ETER, the number includes personnel involved in teaching and research, while in IPEDS, the number of instructional, research and public service staff is used as the nearest equivalent. In both cases, it excludes management, technical and support staff, as well as healthcare staff in hospitals annexed to universities. Coverage of PhD students and postgraduate staff may be incomplete.
- The *total enrolments* at levels 5 (diploma), 6 (bachelor), 7 (master) and 8 (PhD) of the International Standard Classification of Educational Degrees (ISCED).
- The *highest degree* the HEI has the legal right to deliver: diploma, bachelor, master and PhD.

- The *legal status*: institutions under public control or private. In Europe, public institutions also include a small number of HEIs managed by private foundations, but subject to the same rules and funded at the same level as public HEIs, like KU Leuven.
- The *region*: Europe or the US.
- The *number of publications* in Web of Science for the period 2011-2014 (core publications used for the Leiden Ranking only) using fractional counting (the same methodology that is used for the Leiden ranking). For cases not identified in WoS, this indicator was set to zero, but a dummy variable (WoS presence) is introduced to distinguish those HEIs that were identified in the Leiden Ranking.
- The *mean normalized citation score* (MNCS) for publications 2011-2014 as a measure of HEI quality. The indicator uses the same methodology as the Leiden ranking (i.e. citations are counted until the end of 2015). For HEIs not identified in WoS, this indicator was set to zero.

Data availability is high for all indicators, the only exception being academic staff data that are missing for about 20% of the European sample: all HEIs in Austria, Estonia, Latvia, France and Greece. These countries are therefore excluded from the regressions including academic staff.

## 4.2 ResearchGate data

RG data was retrieved from institutional pages (https://www.researchgate.net/institutions/). The HEI list was matched with RG as follows. First, the list of institutional pages in RG was searched automatically by using the institutional name (in English and the national language) and the location to control for similar names. This first search yielded about 4,100 matches. Follow-up checks revealed matching problems due to minor variations between official names and RG names, so manual searches were used to check for different versions of institution names, adding about 300 matches for a total of 4,451 matched records. US multi-campus HEIs (according to IPEDS) with single RG accounts were then aggregated (for example by summing the number of students and of publications), yielding 3,736 matched records (each corresponding to a unique RG URL). It is likely that a few matches were missed, for example due to recent name changes.

### 4.2.1 RG variables

The following variables were derived from ResearchGate.

- RG presence: whether a HEI has an institutional profile on RG.
- RG members: the number of individuals subscribed to RG by institution. Multiple affiliations are not allowed in RG.
- RG publications: the number of publications attributed to a HEI in RG from its members.
- Institutional RG Score: as reported by RG on the institutional profile page[4].

All ResearchGate data were downloaded from June to August 2017. Ratios between these indicators revealed a few cases where the attribution of items to institutions seemed to be incorrect (see also Orduña-Malea and Alonso-Arroyo, 2017 for similar issues with RG institutional profiles of companies). There were five cases with more than 100 RG publications per RG member (the sample mean being 2.6), the most extreme being William Penn University with 82,000 publications from only 32 members (and less than 100 staff FTEs), probably due to confusion with Pennsylvania State University. These five cases were excluded, in addition to the University of Minnesota-Duluth, which has 90,000 RG publications from 1,000 RG members and less than 600 staff FTEs. There were no large additional outliers for RG Score by RG members.

There were 16 cases with more than 10 RG members per staff FTE (the sample mean being 1.57). Most were colleges that were likely to have many part-time teachers; the only large case was Walden University, an on-line university that has 1,000 staff FTEs and more than 11,000 RG members (but only 121 publications on RG). The only case that pointed to data issues was the RG profile of the State of University of New York (SUNY) with 400 members and 18,000 publications. Its IPEDS record refers only to SUNY's administration, while staff and WoS publications are broken down by campus. A manual check showed this profile to contain academics from different universities and so it was excluded. This false attribution presumably resulted in a small reduction of RG values for other SUNY campuses. A few other system administration records for US states were also excluded.

---

[4] https://www.researchgate.net/RGScore/FAQ

### 4.3 Analyses

Besides descriptive analysis, different statistical techniques were employed depending on the characteristics of the dependent variables. Since RG presence is a binary variable, logistic regression is the most suitable choice. Institutional RG scores are counts, so the use of Poisson or Negative Binomial regressions could be suitable for this (Cameron and Trivedi, 1998). However, a log transform strongly reduces the skewedness and kurtosis of the dependent variable, so that an OLS regression can be used, giving a more efficient estimator. Since this approach drops all cases with RG scores equal to zero, a Heckman two-stage regression was performed to test for sampling biases.

Finally, given the strong correlation between academic staff, publication and citations, mediation models were used to disentangle the interaction effects of these variables.

In all cases, the volume variables, such as academic staff, were log transformed and collinearity was tested for by computing Variance Inflation Factors (VIF; O'brien, 2007).

## 5 Results

All variables are highly skewed (Table 1). When excluding zeros, a logarithmic transformation strongly reduced skewness and kurtosis to more acceptable values for OLS regressions – from 4.65 to -0.03 (skewness) and from 31.8 to 2.69 (Kurtosis) for academic staff, which is the main independent variable in the regressions, from 4.62 to -0.33 (skewness) and from 29.3 to 3.50 (kurtosis) for the institutional RG score.

There are 75% more RG members than academic staff, but the latter are in FTEs, while RG could also include profiles of individuals not counted in academic staff, including administrative staff, PhD students and undergraduates[5]. These figures are compatible with the assumption that most academics in European and US universities have RG profiles (according to a 2014 survey, about 88% of scientists were at least aware of it: Van Noorden, 2014), although PhD students are usually not considered to be staff in some institutions, and an unknown proportion of non-research students also have profiles.

There are five times more RG publications than WoS publications. There are three potential sources of this difference, but their relative contributions are unknown: RG publications can be from any year, whereas the WoS publications used here are from 2011-2014; multiple RG authors could upload the same publication on different profiles, while WoS data are based on fractional counting; RG contains many document types that are not in WoS.

Some of these differences may also be generated by the time difference between institutional data (from 2013/4) and RG data (from 2017). Assuming an average annual growth rate for academic staff and WoS publications of 3% (Bornmann and Mutz, 2015), the difference in ratios would be 13%. Regressions are unlikely to be affected by this difference, since the cross-sectional variation is much larger.

---

[5] The inclusion of PhD students among academic staff is a complex issue in HE data (Bonaccorsi *et al*, 2007). In principle, if these students have a contract with the university they should be included, but it is possible that some of them are not counted, particularly those with low employment and those paid by national grants. This will hardly affect the FTEs of academic staff, but might inflate the number of RG members per institution. However, it is unlikely that these individuals have high RG scores and, therefore, the impact on the institutional RG scores is not expected to be very large.

*Table 1. An overview of the data set of universities in Europe and the US: Descriptive statistics*

|  | Mean | STDEV | Min | 1Q | Median | 3Q | Max | Sum | Valid N |
|---|---|---|---|---|---|---|---|---|---|
| Academic staff | 337.52 | 692.23 | 0.00 | 32.00 | 111.00 | 297.00 | 9597.00 | 2026475.79 | 6004.00 |
| Total student enrolments | 5822.23 | 10409.88 | 0.00 | 517.00 | 1933.00 | 6613.00 | 287066.00 | 37733875.00 | 6481.00 |
| Students/staff | 21.84 | 30.94 | 0.00 | 12.75 | 18.13 | 26.33 | 1822.00 | 129153.56 | 5914.00 |
| Wos publications | 298.12 | 1289.50 | 0.00 | 0.00 | 0.00 | 0.00 | 32253.86 | 1971475.82 | 6613.00 |
| WoS publications/staff | 0.18 | 0.55 | 0.00 | 0.00 | 0.00 | 0.00 | 7.71 | 1048.87 | 5935.00 |
| Mean Normalized Citation Score (MNCS) | 0.18 | 0.39 | 0.00 | 0.00 | 0.00 | 0.00 | 4.49 | 1160.28 | 6613.00 |
| RG members | 941.22 | 1930.40 | 1.00 | 58.50 | 202.00 | 769.50 | 20854.00 | 3516389.00 | 3736.00 |
| RG publications | 3181.68 | 10536.27 | 0.00 | 9.00 | 67.00 | 759.50 | 149153.00 | 11886751.00 | 3736.00 |
| Institutional RG score | 5323.67 | 14669.28 | 0.00 | 56.27 | 307.84 | 2107.07 | 159396.90 | 19889239.80 | 3736.00 |
|  |  |  |  |  |  |  |  |  |  |
| Region | US | 4355.00 | Europe | 2258.00 |  |  |  |  |  |
| Highest degree delivered | Diploma | 1543.00 | Bachelor | 1058.00 | Master | 1827.00 | PhD | 2112.00 |  |
| Legal status | Public | 3331.00 | Private | 3280.00 |  |  |  |  |  |
| WoS presence | Yes | 1260.00 | No | 5353.00 |  |  |  |  |  |
| RG presence | Yes | 3736.00 | No | 2877.00 |  |  |  |  |  |

The three RG indicators (RG members, RG publications and institutional RG Score) correlate highly (0.850-0.948) with each other and with academic staff FTEs and WoS publications (Table 2). Correlations between RG indicators and the number of students are lower, even though students and staff FTEs correlate highly (0.814). Thus, staff size and the number of WoS publications are important for RG Score.

The size normalised research quality indicator MNCS has a high correlation (0.988) with publications per staff FTE, which is a research productivity indicator. It also has a strong correlation (0.623) with staff FTEs, confirming that larger institutions tend to have a higher citation impact. The correlation between RG score and MNCS (0.704) is large but lower than with academic staff (0.851) and WoS publications (0.954), suggesting that this RG indicator is more related to volume than to quality.

*Table 2. Spearman correlations between the institutional and RG variables for European and US universities*

|  | Academic staff | Total student enrolments | Students/staff | Wos publications | WoS publications/staff | MNCS | RG members | RG publications | Inst. RG score |
|---|---|---|---|---|---|---|---|---|---|
| Academic staff | 1.000 |  |  |  |  |  |  |  |  |
| Total student enrolments | 0.814 | 1.000 |  |  |  |  |  |  |  |
| Students/staff | -0.384 | 0.138 | 1.000 |  |  |  |  |  |  |
| Wos publications | 0.859 | 0.634 | -0.456 | 1.000 |  |  |  |  |  |
| WoS publications/staff | 0.580 | 0.367 | -0.420 | 0.897 | 1.000 |  |  |  |  |
| Mean Normalized Citation Score (MNCS) | 0.632 | 0.517 | -0.224 | 0.986 | 0.988 | 1.000 |  |  |  |
| RG members | 0.821 | 0.716 | -0.263 | 0.852 | 0.679 | 0.597 | 1.000 |  |  |
| RG publications | 0.802 | 0.584 | -0.432 | 0.935 | 0.844 | 0.659 | 0.850 | 1.000 |  |
| Institutional RG score | 0.851 | 0.642 | -0.420 | 0.954 | 0.836 | 0.704 | 0.908 | 0.948 | 1.000 |
| All correlations are significant at the 0.001 level (2-tailed). | | | | | | | | | |

## 5.1 HEI presence on RG

Among the 6,613 HEIs in the dataset, 3,736 (56%) had an institutional RG profile. While 83% of PhD-awarding HEIs have an institutional profile in RG, the share is 57% for master, 33% for bachelor, and 37% for diploma (Figure 1).

European HEIs are more present in RG than their US counterparts, but this difference is due to the inclusion of associate colleges in the US. When comparing only HEIs delivering at least a bachelor degree, US universities are more present in RG than their European counterparts. Public HEIs are also more frequently present in RG than private HEIs, but this difference is affected by private HEIs delivering lower degrees and being less research oriented.

*Figure 1. Institutional profile in ResearchGate by group of HEIs.*

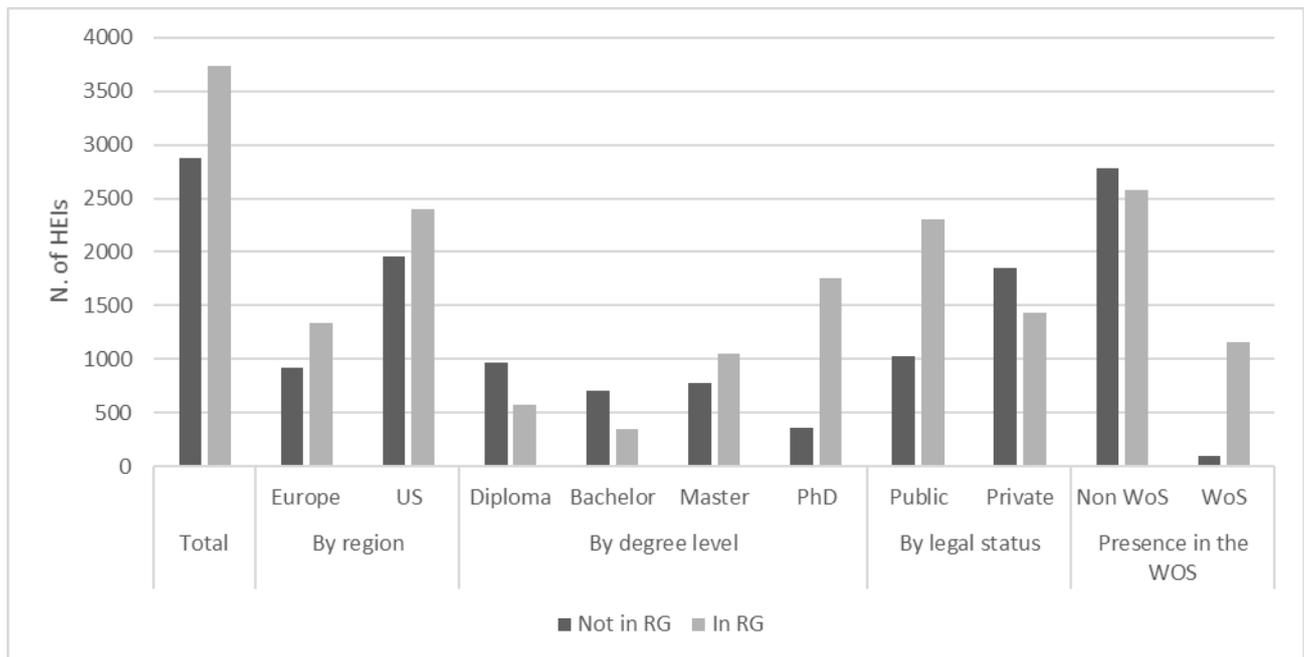

Out of 1,260 HEIs found in WoS, only 101 did not have an institutional profile in RG. A further check revealed that most of these cases had an institutional profile that had not been found. About half of the HEIs not found in WoS have a RG profile. This paper does not investigate the kinds of publications in RG for HEIs not found in WoS. It is possible that some of these have WoS publications from other years, or publications that were not found in the Leiden ranking or members from these HEIs are publishing in RG non-WoS publications.

A logistic regression was run for RG presence against HEI characteristics to untangle the importance of the independent variables. The regression uses the log of the number of academic staff as an HEI size variable, students by staff and publications by staff to reflect educational intensity and research productivity respectively, and MNCS as an average citation impact indicator. These variables have weak or moderate correlations (maximum Pearson correlation: 0.584), except for MNCS and publications/staff (correlation: 0.774). Three categorical variables were introduced for the highest degree delivered, legal status, and region (Europe or US).

*Table 3. Four logistic regressions for institutional presence on RG.*

|  | Staff only | | | Staff and students | | | + Categorial variables | | | + Region | | |
|---|---|---|---|---|---|---|---|---|---|---|---|---|
|  | B | SE | Sig. | B | SE | Sig. | B | SE | Sig. | B | SE | Sig. |
| Constant | -5.452 | 0.151 | 0.000 | -5.414 | 0.165 | 0.000 | -7.060 | 0.229 | 0.000 | -8.569 | 0.263 | 0.000 |
| log academic staff | 1.281 | 0.033 | 0.000 | 1.288 | 0.036 | 0.000 | 1.424 | 0.044 | 0.000 | 1.385 | 0.045 | 0.000 |
| student / staff |  |  |  | -0.005 | 0.002 | 0.047 | 0.002 | 0.001 | 0.233 | 0.002 | 0.001 | 0.138 |
| publications / staff |  |  |  | 0.052 | 0.185 | 0.779 | 0.024 | 0.178 | 0.894 | 0.019 | 0.191 | 0.920 |
| mncs |  |  |  | 0.416 | 0.207 | 0.044 | -0.261 | 0.210 | 0.215 | 0.163 | 0.225 | 0.469 |
| Degree = bachelor |  |  |  |  |  |  | 0.336 | 0.113 | 0.003 | 0.980 | 0.127 | 0.000 |
| Degree = master |  |  |  |  |  |  | 0.878 | 0.098 | 0.000 | 2.081 | 0.131 | 0.000 |
| Degree = PhD |  |  |  |  |  |  | 1.378 | 0.122 | 0.000 | 2.406 | 0.144 | 0.000 |
| Legal status = Private |  |  |  |  |  |  | 0.588 | 0.092 | 0.000 | -0.111 | 0.107 | 0.297 |
| Region = US |  |  |  |  |  |  |  |  |  | 1.812 | 0.113 | 0.000 |
| N | 5935 | | | 5914 | | | 5914 | | | 5914 | | |
| AIC | 5249.094 | | | 5152.668 | | | 4865.680 | | | 4576.093 | | |
| % correctly classified | 80% | | | 80% | | | 82% | | | 83% | | |

The basic model with only academic staff classifies 80% of the cases correctly (against 57% for the null model). Including covariates for education and research does not greatly improve model fit and the percentage of correctly classified cases, probably because the different HEI characteristics are strongly correlated (larger HEIs are also more research oriented). The variables for the degrees awarded confirm that the likelihood to be in RG increases with the higher level of degrees delivered, even when controlling for size. Private HEIs have a higher likelihood to be in RG when controlling for size and degree level, but the effect becomes non-significant when controlling for the region, so this factor seems to be irrelevant. In contrast, US HEIs were more present in RG after accounting for other institutional characteristics.

Summarising the final model, the key factors associated with RG presence are institutional size (academic staff FTEs), level of degree awarded (the higher the better), and US location. Education orientation (students/staff), research productivity (publications/staff), research quality (MNCS), and legal status are not relevant.

These results are better analysed in terms of the number of staff for which the expected probability of having an institutional profile is 0.5. This threshold is at 176 FTE of academic staff in Europe and 47 FTEs in the US for PhD-awarding HEIs, but increases to 493 FTEs and to 133 FTEs for HEIs awarding bachelor degrees in Europe and in the US respectively. In substantive terms, this means that practically all PhD-awarding HEIs are in RG, regardless of their size, while only the larger colleges awarding degrees at the diploma and bachelor levels are present in RG, consistent with the descriptive statistics.

The results were analysed by the presence and absence in one or both of RG and WoS (Figure 2). HEIs without an RG institutional profile form 42% of the sample but account for a small share of academic staff FTEs and students. In contrast, the 2,577 HEIs not found in WoS but present in RG account for 27% of academic staff FTEs and 36% of the enrolled students, but for only 6% of RG publications and 8% of total institutional RG Scores. Most of these are from a few universities that were not correctly matched in WoS (particularly some medical research centres), while the remaining HEIs in this group have few publications in RG. Half of the HEIs without an RG institutional profile award master's degrees, and half award degrees at the bachelor and diploma levels.

HEIs with WoS publications mostly award PhDs. These 1,159 HEIs (17% of the sample) constitute one-third of the HEIs in RG, but account for over 90% of the RG publications and institutional RG Scores (as compared with 50% of academic staff).

In summary, while large HEIs without many publications in WoS tend to be in RG, the volume of their presence is low – supporting the previous results that the volume of RG presence correlates with publishing in WoS-indexed outlets.

*Figure 2. Subgroups of HEIs by RG presence and publishing status*

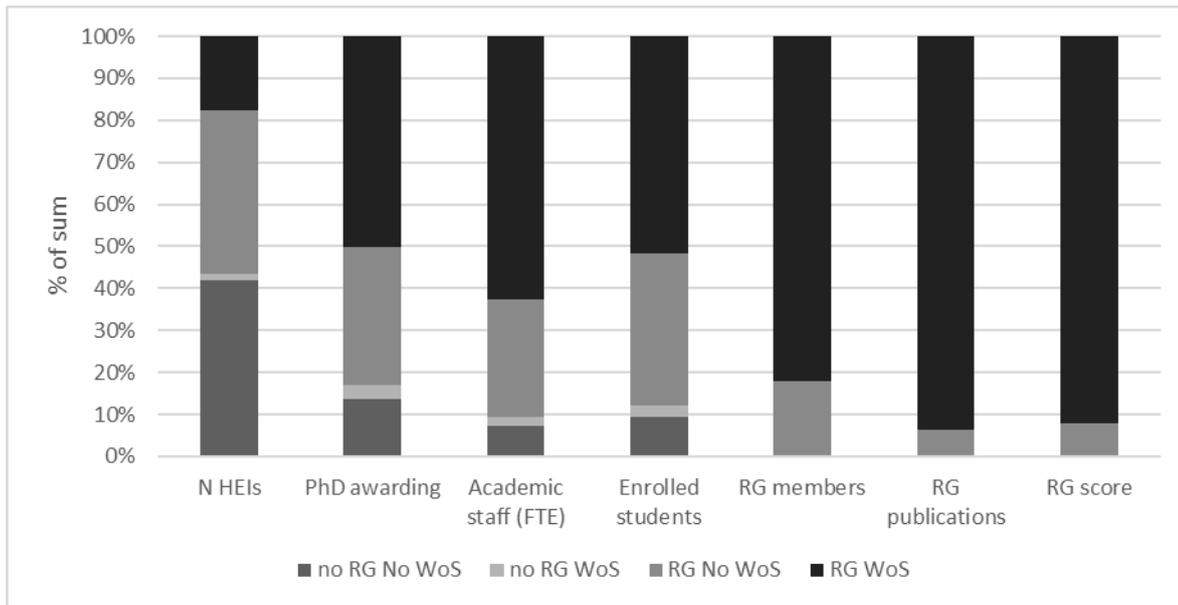

## 5.2 Institutional RG Score

An OLS regression on the HEIs present in RG was conducted to investigate factors associating with higher institutional RG Scores. A Heckman selection model on the full sample using ln(staff) as the selection variable provides similar results. The RG score was log-transformed to reduce skewness. The controls for the model were students per staff FTE, PhD-awarding, region (US or Europe) and legal status (private or public). Correlations between these variables are low or moderate except for the one between publications per staff FTE and MNCS (Pearson correlation coefficient: 0.774).

The first regression uses all HEIs in RG to determine whether academic staff (log-transformed) and publications per staff FTE associate with institutional RG Scores. This sample also includes HEIs that do not regularly publish. Academic staff FTEs is a highly significant covariate ($p<0.001$). The number of students per staff FTE (the educational orientation indicator) is positive, but only marginally significant, while no significant difference between US and European HEIs are found. Furthermore, private institutions have lower RG scores than their public counterparts. Other factors being the same, delivering higher degrees consistently increases RG scores.

Introducing publications per staff FTE only slightly increases the model fit; this may be expected since two-thirds of the HEIs in this sample have no RG publications. The quality indicator MNCS has even less of an impact on the model fit. The Variance Inflation Factors (VIF) are low enough to avoid concern, as a score of 10 is sometimes used as the minimum VIF to be problematic (O'brien, 2007). The introduction of MNCS also only moderately affects the publications per staff coefficient, suggesting that collinearity is not a major issue.

Because the dependent variable is log transformed, regression coefficients should be interpreted as multiplicative factors. For example, an increase of 0.55 in publications per staff (i.e. a standard deviation of this variable) multiplies the RG Score of an HEI by exp(0.607*0. 55)= 1.4. The right to award a PhD multiplies the RG Score by exp(0.582)=1.79 with respect to an HEI delivering master degrees (i.e. with the same covariates, a PhD awarding HEI will have a RG score 80% larger than a HEI that only awards masters diplomas).

*Table 4. Results from three OLS regressions for log RG Score for HEIs with an institutional profile in RG*

|  | Size and inst. charact. | | | Including publications | | | Including impact | | | |
|---|---|---|---|---|---|---|---|---|---|---|
|  | Coef. | Std. Err. | P>t | Coef. | Std. Err. | P>t | Coef. | Std. Err. | P>t | VIF |
| Constant | -4.363 | 0.181 | 0.000 | -3.497 | 0.179 | 0.000 | -1.095 | 0.185 | 0.000 | |
| Log_staff | 1.380 | 0.026 | 0.000 | 1.189 | 0.027 | 0.000 | 1.144 | 0.028 | 0.000 | 2.13 |
| Publications per staff | | | | 0.798 | 0.044 | 0.000 | 0.607 | 0.055 | 0.000 | 2.52 |
| Students per staff | 0.000 | 0.002 | 0.872 | 0.004 | 0.002 | 0.026 | 0.004 | 0.002 | 0.037 | 1.21 |
| MNCS | | | | | | | 0.551 | 0.097 | 0.000 | 3.37 |
| Degree = bachelor | 2.600 | 0.114 | 0.000 | 2.529 | 0.109 | 0.000 | 2.499 | 0.108 | 0.000 | 1.80 |
| Degree = master | 2.653 | 0.095 | 0.000 | 2.621 | 0.091 | 0.000 | 2.602 | 0.091 | 0.000 | 2.85 |
| Degree = PhD | 3.588 | 0.096 | 0.000 | 3.300 | 0.093 | 0.000 | 3.184 | 0.095 | 0.000 | 3.84 |
| Region = US | -0.028 | 0.067 | 0.677 | 0.001 | 0.064 | 0.983 | 0.075 | 0.065 | 0.246 | 1.55 |
| Legal status = Private | -0.318 | 0.071 | 0.000 | -0.307 | 0.068 | 0.000 | -0.310 | 0.067 | 0.000 | 1.88 |
| N | 3410 | | | 3410 | | | 3410 | | | |
| AIC | 12355.54 | | | 12043.57 | | | 12013.18 | | | |
| Rsquared | 0.721 | | | 0.746 | | | 0.748 | | | |

The results change when only WoS-publishing HEIs are considered (Table 5). The model without WoS publications provides a similar fit to that of the full sample (Table 5) but, for the reduced data set, introducing the number of publications per staff FTE substantially increases the level of fit (Table 6), as shown by the Rsquare and the Akaike Information Criterion (AIC). This shows that the number of publications on Web of Science (WoS) is an important explanatory variable for the sample of publishing HEIs. Once the number of publications is included, differences between the US and Europe are no longer significant. Lastly, after introducing the quality of the HEIs proxied by MNCS, the increase in the model fit and the drop in AIC are small. Thus, HEI quality has little impact on RG Score, particularly when compared with the number of publications. Finally, private HEIs have lower scores than public ones.

The most important result however is that the alternative model that includes only the logged number of WoS publications (instead of logged academic staff and publications per staff) provided the best overall fit, implying that the simple count of WoS publications is statistically the best predictor of RG score. With the exception of legal status, all other covariates are no longer statistically significant.

*Table 5. Results from three OLS regressions for log RG Score for HEIs with an institutional profile in RG and WoS publications*

*Only the PhD awarding dummy is included, as there are very few HEIs in the WoS delivering only short degrees or masters.*

|  | Size and inst. charact. | | | Including publications | | | Including impact | | | | Publications only | | |
|---|---|---|---|---|---|---|---|---|---|---|---|---|---|
|  | Coef. | Std. Err. | P>t | Coef. | Std. Err. | P>t | Coef. | Std. Err. | P>t | VIF | Coef. | Std. Err. | P>t |
| Constant | -0.554 | 0.250 | 0.027 | 0.368 | 0.205 | 0.073 | 0.246 | 0.207 | 0.234 | | 0.2461 | 0.207 | 0.234 |
| Log_staff | 1.240 | 0.034 | 0.000 | 1.025 | 0.029 | 0.000 | 1.005 | 0.029 | 0.000 | 1.530 | | | |
| Log_publications | | | | | | | | | | | 0.7389 | 0.013 | 0.000 |
| Publications per staff | | | | 0.689 | 0.030 | 0.000 | 0.637 | 0.033 | 0.000 | 1.650 | | | |
| Students per staff | -0.015 | 0.003 | 0.000 | -0.004 | 0.002 | 0.084 | -0.004 | 0.002 | 0.075 | 1.150 | -0.0008 | 0.002 | 0.698 |
| MNCS | | | | | | | 0.337 | 0.092 | 0.000 | 1.450 | 0.1355 | 0.075 | 0.073 |
| PhD awarding | 1.294 | 0.117 | 0.000 | 0.969 | 0.096 | 0.000 | 0.972 | 0.095 | 0.000 | 1.120 | 0.0714 | 0.086 | 0.404 |
| Region = US | 0.167 | 0.069 | 0.015 | -0.019 | 0.056 | 0.732 | -0.020 | 0.056 | 0.715 | 1.250 | 0.0093 | 0.047 | 0.844 |
| Legal status = Private | -0.550 | 0.088 | 0.000 | -0.502 | 0.071 | 0.000 | -0.521 | 0.070 | 0.000 | 1.380 | -0.2487 | 0.061 | 0.000 |
| N | 1008 | | | 1008 | | | 1008 | | | | 1008 | | |
| AIC | 2788.856 | | | 2355.861 | | | 2344.272 | | | | 2007.102 | | |
| Rsquared | 0.719 | | | 0.817 | | | 0.819 | | | | 0.871 | | |

Given that all these variables are strongly correlated, and specifically that academic staff and publications cannot be introduced together in a regression for log(RG score) for WoS-publishing institutions, a mediation model was used to assess the relative importance of the different paths for RG Score. In the first model (Figure 3), the number of academic staff influenced the RG score both directly and through the number of publications. The second model also incorporates the effect of quality through MNCS.

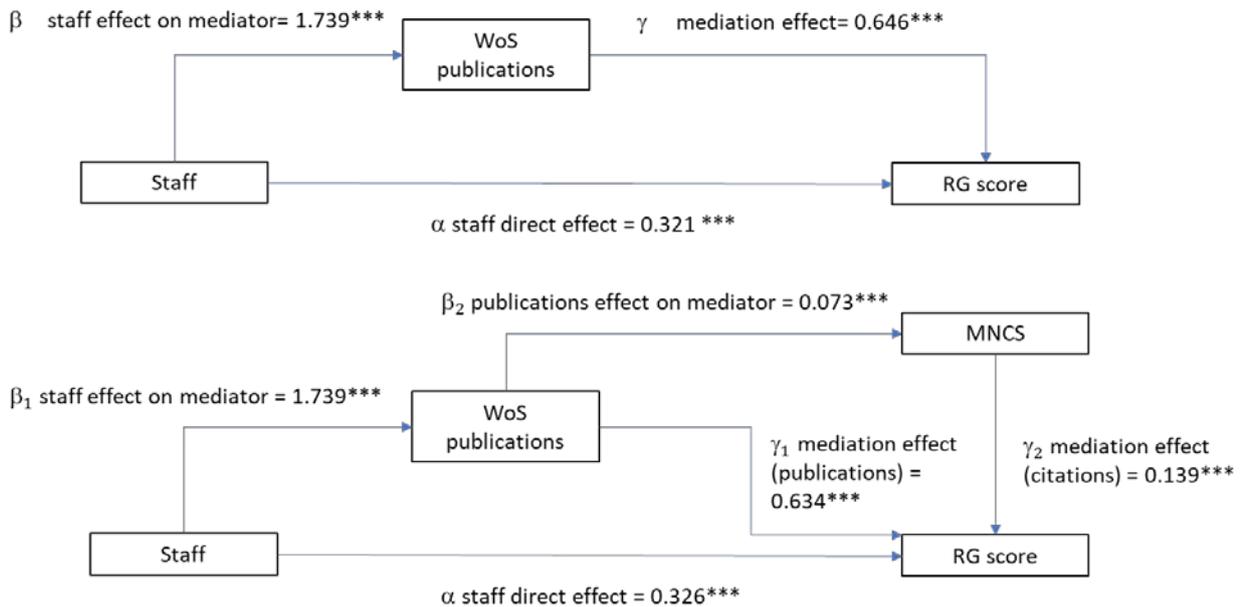

*Figure 3. Mediation model to assess the influence of factors affecting RG Score for institutions in both WoS and RG.*

*Significance codes: *<0.05, ***<0.001.*

From the first model, while the number of academic staff has an association with RG Score independent from the number of WoS publications, the direct effect (0.321) is much smaller than the indirect one (0.646*1.739 = 1.123). The second model incorporates the effect of quality. Whilst citations counts are associated with the RG Score, the effect through MNCS (0.139*0.073*1.739=0.017) is much smaller than the effect through publications (0.634*1.739=1.103). Overall, the direct effect of staff accounts for 23% of the variance in the RG Score, while publications accounted for 76% and publications and citations combined for 1%. In other words, while institutional RG Scores reflect a mix of size, publication output and quality – with all these characteristics being correlated – the path going through the number of WoS publications is the dominant one in accounting for differences between HEIs publishing in WoS in their RG Scores.

### 5.3 Limitations

The analysis is limited by human errors in the data collection, the coverage of only two geographic regions, and the data collection dates. The RG Score algorithm may change in the future, as well as the patterns of joining the site. The independent variables are not fully independent of each other due to unmodeled factors, such as legislation and national economic policies that affect universities in groups (all those in one country) rather than individually. The dichotomy between public and private universities hides the existence of multiple types of private universities, including religious schools.

A limitation for the RG Scores regression is that the publication component of the RG Score is likely to use whole counting (i.e., each author is treated as the sole author of all their papers), whereas the Leiden publication data uses fractional counting (i.e., if there are n authors then each author is treated as having written 1/n of the paper) that is generally preferred in bibliometrics. However, the aggregate difference between the two indicators is very small (e.g., a correlation of 0.97 between the two approaches for a percentile indicator; Waltman *et al*, 2012), so this limitation is unlikely to affect the results.

## 6  Discussion and conclusions

The results show that large numbers of institutions of all types have joined ResearchGate, including those that do not award PhDs and those that do not publish in Web of Science. Being present in WoS is almost a sufficient condition for having an institutional profile in RG, demonstrating that RG at least covers the world of HEIs publishing in conventional outlets.

Not all institutions have a presence in RG. Based on the regression results, larger institutions (academic staff FTEs), higher levels of degree awarding powers, and geographic location in the US all make an institution more likely to have an RG presence. In contrast, educational orientation (students/staff FTEs), research productivity (publications/staff FTEs), citation impact (MNCS), and legal status do not seem relevant. Despite the multiple statistically significant regression coefficients, institutional size alone is a good predictor of membership (80% correct from a baseline 57%) and the other variables only increase the prediction rate by 3%, so institution size seems to be the dominant factor for RG membership.

From the regression for all HEIs in RG, the main explanatory factor for institutional RG score is represented by institutional size (log staff FTEs). This is reasonable since two-thirds of these HEIs had no publications in WoS. When considering only WoS publishing HEIs, publication activity becomes an important factor for RG scores, while WoS normalized impact contributes less. Even though institutional size, publication volume and quality are all correlated, the model shows that WoS publication volume is statistically the main single explanatory factor for RG Score. After accounting for it, differences between Europe and the US also disappear. Finally, public HEIs have higher scores than their private counterparts, something that may be explained by the fact that private HEIs can be more focused on students and economics than on research.

While the mechanisms generating the patterns observed are not directly tested by the regression approach, some assumptions can still be made. First, if each European or US academics had a similar probability to join RG, a strong correlation with the number of academic staff would be expected. Moreover, it seems likely that non-publishing academics are less likely to join RG, so non-WoS HEIs would have a lower percentage of members on RG, meaning larger HEIs are more likely to be present in RG even if they are not present in the WoS. Second, if individual level activity is more essential for earning high RG Scores on the site when compared to achievements, (Orduna-Malea et al, 2017), the institutional RG Score is likely to reflect aggregated members' activities within RG. Moreover, while individual members may generate activity, for example by uploading publications or providing answers to questions, RG also populates members' profiles with publications retrieved automatically from academic journals. This would help to explain why the aggregated activity at the HEI level so strongly correlates with (scholarly) publication counts, as measured by the number of WoS publications.

Beyond statistical regularities, these findings have important implications for the use of RG indicators. First, institutional presence and institutional RG Scores should not be interpreted as strong indicators of research impact or productivity, due to the importance of institutional size. RG Scores are so highly correlated with the volume of WoS publications that they should not be used for comparisons or for ranking institutions for research quality, because larger institutions will tend to have an unjustifiably high research performance ranking. For example, a large weak organisation may be ranked higher than a small but excellent unit. Second, at the institutional level, the academic world, as represented by RG indicators, is not significantly different from the one depicted by more conventional bibliometric indicators, therefore questioning their added value for institutional comparisons. Third, these findings may be due in large part to the strategy adopted by RG of searching for potential users and populating members' accounts from databases like the WoS automatically and providing scores that may reflect the legitimate commercial interests of RG rather than a goal to provide robust and useful indicators for the academic community (Copiello and Bonifaci, 2018).

# 7 Acknowledgments

This research was supported by the European Union through the project, Research infrastructures for the assessment of science, technology and innovation policy (RISIS) (Grant agreement no: 313082). Thank you to the European Commission, the ETER project, and to CWTS, University of Leiden, for providing access to the data.

# 8 References

Alheyasat, Omar 2015. Examination expertise sharing in academic social networks using graphs: The case of ResearchGate. *Contemporary Engineering Sciences,* 8, 137-151.

Ali, Muhammad Yousuf, Malcolm Wolski and Joanna Richardson 2017. Strategies for using ResearchGate to improve institutional research outcomes. *Library Review,* 66, 726-739.

Baro, Ebikabowei Emmanuel and Monica Eberechukwu Eze 2017. Perceptions, Preferences of Scholarly Publishing in Open Access Routes: A Survey of Academic Librarians in Nigeria. *Information and Learning Science,* 118,

Bonaccorsi, Andrea, Cinzia Daraio, Benedetto Lepori and Stig Slipersaeter 2007. Indicators on individual higher education institutions: addressing data problems and comparability issues. *Research Evaluation,* 16, 66-78.

Bornmann, Lutz and Rüdiger Mutz 2015. Growth rates of modern science: A bibliometric analysis based on the number of publications and cited references. *Journal of the Association for Information Science and Technology,* 66, 2215-2222.

Cameron, A. C. and P. K. Trivedi. 1998. Regression analysis of count data. Econometric Society Monograph, Cambridge University Press.

Chakraborty, Nirmali 2012. Activities and reasons for using social networking sites by research scholars in NEHU: A study on Facebook and ResearchGate.

Chawla, Dalmeet Singh. 2017. Publishers take academic networking site to court. American Association for the Advancement of Science.

Copiello, Sergio and Pietro Bonifaci 2018. A few remarks on ResearchGate score and academic reputation. *Scientometrics,* 1-6.

Goodwin, Spencer, Wei Jeng and Daqing He 2014. Changing communication on ResearchGate through interface updates. *Proceedings of the Association for Information Science and Technology,* 51, 1-4.

Hoffmann, Christian Pieter, Christoph Lutz and Miriam Meckel 2016. A relational altmetric? Network centrality on ResearchGate as an indicator of scientific impact. *Journal of the Association for Information Science and Technology,* 67, 765-775.

Iglesias-García, Mar, Cristina González-Díaz and Lluís Codina 2017. A Study of Student and University Teaching Staff Presence on ResearchGate and Academia. edu in Spain. In *Media and Metamedia Management,* ed. Anonymous , pp. 509-515. Springer.

Jamali, Hamid R. and Majid Nabavi 2015. Open access and sources of full-text articles in Google Scholar in different subject fields. *Scientometrics,* 105, 1635-1651.

Jamali, Hamid R., David Nicholas and Eti Herman 2015. Scholarly reputation in the digital age and the role of emerging platforms and mechanisms. *Research Evaluation,* 25, 37-49.

Jeng, Wei, Spencer DesAutels, Daqing He and Lei Li 2017. Information exchange on an academic social networking site: A multidiscipline comparison on researchgate Q&A. *Journal of the Association for Information Science and Technology,* 68, 638-652.

Jordan, Katherine 2015. Exploring the ResearchGate score as an academic metric: Reflections and implications for practice.

Jordan, Katy. 2017. Understanding the structure and role of academics' ego-networks on social networking sites. The Open University.

Kraker, Peter and Elisabeth Lex. 2015. A critical look at the ResearchGate score as a measure of scientific reputation.

Kuo, Tsuang, Gwo Yang Tsai, Yen-Chun Jim Wu and Wadee Alhalabi 2016. From sociability to creditability for academics. *Comput.Hum.Behav.,*

Laakso, Mikael, Juho Lindman, Cenyu Shen, Linus Nyman and Bo-Christer Björk 2017. Research output availability on academic social networks: implications for stakeholders in academic publishing. *Electronic Markets,* 27, 125-133.

Li, Lei, Daqing He, Wei Jeng, Spencer Goodwin and Chengzhi Zhang. 2015. Answer quality characteristics and prediction on an academic Q&A Site: A case study on ResearchGate. ACM.

Madhusudhan, Margam 2012. Use of social networking sites by research scholars of the University of Delhi: A study. *The International Information & Library Review,* 44, 100-113.

Manca, Stefania and Maria Ranieri 2017. Implications of social network sites for teaching and learning. Where we are and where we want to go. *Education and Information Technologies,* 22, 605-622.

Martín-Martín, Alberto, Enrique Orduna-Malea and Emilio Delgado López-Cózar 2018. Author-level metrics in the new academic profile platforms: The online behaviour of the Bibliometrics community. *Journal of Informetrics,* 12, 494-509.

Martín-Martín, Alberto, Enrique Orduña-Malea, Juan M. Ayllón and Emilio Delgado López-Cózar 2016. The counting house: Measuring those who count. Presence of bibliometrics, scientometrics, informetrics, webometrics and altmetrics in the Google Scholar citations, Researcherid, ResearchGate, Mendeley & Twitter. *arXiv preprint arXiv:1602.02412,*

Mas-Bleda, Amalia, Mike Thelwall, Kayvan Kousha and Isidro F. Aguillo 2014. Do highly cited researchers successfully use the social web? *Scientometrics,* 101, 337-356.

Meishar-Tal, Hagit and Efrat Pieterse 2017. Why Do Academics Use Academic Social Networking Sites? *The International Review of Research in Open and Distributed Learning,* 18,

Mikki, Susanne, Marta Zygmuntowska, Øyvind Liland Gjesdal and Hemed Ali Al Ruwehy 2015. Digital Presence of Norwegian Scholars on Academic Network Sites—Where and Who Are They? *PloS one,* 10, e0142709.

Muscanell, Nicole, Nicole Muscanell, Sonja Utz and Sonja Utz 2017. Social networking for scientists: an analysis on how and why academics use ResearchGate. *Online Information Review,* 41, 744-759.

O'brien, Robert M. 2007. A caution regarding rules of thumb for variance inflation factors. *Quality & Quantity,* 41, 673-690.

Onyancha, Omwoyo B. 2015. Social media and research: an assessment of the coverage of South African universities in ResearchGate, Web of Science and the Webometrics Ranking of World Universities. *South African Journal of Libraries and Information Science,* 81, 8-20.

Orduna-Malea, Enrique, Alberto Martín-Martín, Mike Thelwall and Emilio Delgado López-Cózar 2017. Do ResearchGate Scores create ghost academic reputations? *Scientometrics,* 1-18.

Orduña-Malea, Enrique and Adolfo Alonso-Arroyo. 2017. Cybermetric Techniques to Evaluate Organizations Using Web-based Data. Chandos Publishing.

Ortega, José Luis 2017. Toward a homogenization of academic social sites: A longitudinal study of profiles in Academia. edu, Google Scholar Citations and ResearchGate. *Online Information Review,* 41, 812-825.

Ortega, José Luis 2015. Relationship between altmetric and bibliometric indicators across academic social sites: The case of CSIC's members. *Journal of Informetrics,* 9, 39-49.

Ovadia, Steven 2014. ResearchGate and Academia. edu: Academic social networks. *Behavioral & Social Sciences Librarian,* 33, 165-169.

Stuart, Andrew, Sarah P. Faucette and William Joseph Thomas 2017. Author Impact Metrics in Communication Sciences and Disorder Research. *Journal of Speech, Language, and Hearing Research,* 60, 2704-2724.

Sud, Pardeep and Mike Thelwall 2014. Evaluating altmetrics. *Scientometrics,* 98, 1131-1143.

Thelwall, Mike and Kayvan Kousha 2017a. ResearchGate articles: Age, discipline, audience size, and impact. *Journal of the Association for Information Science and Technology,* 68, 468-479.

Thelwall, Mike and Kayvan Kousha 2017b. ResearchGate versus Google Scholar: Which finds more early citations? *Scientometrics,* 1-7.

Thelwall, Mike 2016. Interpreting correlations between citation counts and other indicators. *Scientometrics,* 108, 337-347.

Thelwall, Mike and Kayvan Kousha 2015. ResearchGate: Disseminating, communicating, and measuring Scholarship? *Journal of the Association for Information Science and Technology,* 66, 876-889.

Van Noorden, R. 2014. Online collaboration: Scientists and the social network. *Nature,* 512, 126-129.

Waltman, Ludo, Clara Calero-Medina, Joost Kosten, et al 2012. The Leiden Ranking 2011/2012: Data collection, indicators, and interpretation. *J.Am.Soc.Inf.Sci.Technol.,* 63, 2419-2432.

Williams, Ann E. and Melissa A. Woodacre 2016. The possibilities and perils of academic social networking sites. *Online Information Review,* 40, 282-294.

Wilsdon, J., L. Allen, E. Belfiore, et al 2015. The Metric Tide: Report of the Independent Review of the Role of Metrics in Research Assessment and Management. Bristol: Higher Education Funding Council for England.

Yan, Weiwei, Yin Zhang and Wendy Bromfield 2018. Analyzing the follower–followee ratio to determine user characteristics and institutional participation differences among research universities on ResearchGate. *Scientometrics,* 115, 299-316.

Yu, Min-Chun, Yen-Chun Jim Wu, Wadee Alhalabi, Hao-Yun Kao and Wen-Hsiung Wu 2016. ResearchGate: An effective altmetric indicator for active researchers? *Comput.Hum.Behav.,* 55, 1001-1006.